\documentstyle[epsf]{l-aa}

\newcommand{\subs}[1]{\mbox{\scriptsize{#1}}}

\begin{document}

\thesaurus{03 (                        
              11.12.2;                 
              11.17.3;                 
              11.19.1                  
               )}

\title{The local luminosity function of QSOs and Seyfert~1 nuclei%
   \thanks{Based on observations made at the European Southern Observatory, 
           La Silla, Chile}
       }

\author{T. K\"ohler \and D. Groote \and D. Reimers \and L. Wisotzki}

\institute{Hamburger Sternwarte, Gojenbergsweg 112,
           D-21029 Hamburg, Germany 
           }

\offprints{L.~Wisotzki, \protect\\ 
           email: lwisotzki@hs.uni-hamburg.de}
\date{Received ; accepted} 
\maketitle
\markboth{T. K\"ohler et al.: Local luminosity function of quasars}{}

\begin{abstract}
We present the analysis of a new flux-limited sample of 
bright quasars and Seyfert~1 galaxies in an 
effective area of 611\,deg$^2$, drawn from the 
Hamburg/ESO survey. 
We confirm recent claims that bright quasars have a
higher surface density than previously thought.
Special care was taken to avoid morphological 
and photometric biases against low-redshift quasars,
and about 50\,\% of the sample objects are at $z<0.3$,
spanning a range of $\Delta M \simeq 8$ in absolute magnitudes.
While our derived space densities for low-luminosity
Seyfert~1 nuclei are consistent with those found in the
literature, we find that luminous QSOs, $M_B < -24$,
are much more numerous in the local universe than
previous surveys indicated.
The optical luminosity functions of 
Seyfert~1 nuclei and QSOs join smoothly, 
and if the host galaxy contributions are taken into account, 
a single power-law of slope $\alpha = -2.2$
describes the combined local luminosity function 
adequately, over the full range in absolute magnitude.
Comparing our data with published results at higher
redshifts, we can rule out pure luminosity evolution
as an acceptable parametrisation; the luminosity
function of quasars changes shape and slope with $z$,
in the sense that the most luminous quasars show the weakest
evolution.

\end{abstract}

\section{Introduction}

The average luminosities of quasars, and possibly many other
properties, depend strongly on cosmological epoch. 
Shape, normalisation, and evolution of the quasar luminosity 
function (QLF) are among the most basic descriptions of the 
quasar population. 
A good knowledge of the {\em local\/} ($z\approx 0$) QLF is 
essential for at least two purposes: first, to serve as zero-point
for quasar evolution studies; second, to provide a reference
distribution law for unbiased 
subsamples constructed for statistical investigations (e.g., for
spectral properties, host galaxies, etc).

Unfortunately, current determinations of the QLF are 
limited mainly to $z\ga 0.3$, for the 
simple reason of lacking appropriate survey data.
Most optical QSO surveys discriminate against objects 
with extended morphology, causing severe incompleteness 
(and bias) at low redshifts. 
At the low-luminosity end of the QSO--Seyfert family,
dedicated galaxy surveys have prompted
investigations of the Seyfert~1 luminosity function 
(e.g., Cheng et al.\ \cite{cheng:85};
Huchra \& Burg \cite{huchra:92}).
However, these surveys are necessarily
incomplete at {\em high\/} nuclear luminosities, where the 
active nuclei outshine the surrounding hosts,
and the objects are not selected as galaxies anymore.

In 1990 we started the `Hamburg/ESO survey' (HES),
a new wide-angle survey for bright
QSOs and Seyferts, designed to avoid the above selection
biases as far as possible. A detailed description of the HES
was given by Wisotzki et al.\ (\cite{wisotzki:96}; hereafter Paper~I),
and a first list of 160 newly discovered QSOs 
was published by Reimers et al.\ (\cite{reimers:96}; Paper~II).
Basically, QSO candidates are selected by automated procedures 
from digitised objective-prism plates
taken with the ESO Schmidt telescope, with subsequent slit
spectroscopy of all candidates above the magnitude limit of
typically $B_{\subs{lim}} \simeq 17.0$--17.5, over an 
area of $\sim 5000$\,deg$^2$ at high galactic latitude.
To prevent morphological bias against sources with resolved
structure, such as Seyferts or gravitational lenses, 
objects with non-stellar appearance 
are {\em not\/} excluded from the candidate lists. 
Among other goals, we aim to construct a large, 
optically flux-limited sample of low-redshift 
`Type~1 AGN', including QSO/Seyfert borderline objects.

In this paper we publish the results of an analysis 
of a first part of the HES, based on 33 Schmidt fields,
with an effective area of 611\,deg$^2$.
We present the first direct determination of
the local quasar luminosity function from a single
optical survey, and discuss some implications of our results.
A comprehensive description of the analysis
has been given by K\"ohler (\cite{koehler:96a}).
We adopt a cosmological model with
$H_0 = 50\,\mathrm{km\,s}^{-1}\,\mathrm{Mpc}^{-1}$,
$q_0 = 0.5$, and $\Lambda=0$.

\begin{table}[tb]
\caption[]{Properties of surveyed fields. Field
designations are given as standard ESO/SERC field
numbers. $N_H$ is
the column density of Galactic neutral hydrogen in
the field centre, in $10^{20}$\,cm$^{-2}$. $A_B$
is the corresponding extinction in the $B$ band.
$\Omega_{\subs{eff}}$ is the effective survey area for
the field, in deg$^2$.}
\label{tab:fields}
\begin{tabular}{lllll}
Field & $B_{\subs{lim}}$ & $N_H$ & $A_B$ & $\Omega_{\subs{eff}}$ \\[0.3ex]
\hline
501 & 17.38 & 5.5 & 0.39 & 16.38 \rule{0em}{2.7ex} \\
503 & 17.25 & 4.8 & 0.34 & 17.75 \\
505 & 16.95 & 7.4 & 0.53 & 17.05 \\
506 & 17.66 & 7.5 & 0.54 & 17.17 \\
507 & 17.11 & 7.2 & 0.51 & 17.10 \\
509 & 16.95 & 5.3 & 0.38 & 17.04 \\
568 & 16.95 & 5.8 & 0.41 & 18.20 \\
570 & 16.78 & 4.2 & 0.30 & 18.92 \\
578 & 16.64 & 6.2 & 0.44 & 17.59 \\
637 & 16.90 & 5.5 & 0.39 & 17.94 \\
638 & 16.95 & 7.5 & 0.53 & 16.14 \\
639 & 17.47 & 5.1 & 0.36 & 19.35 \\
640 & 17.50 & 4.9 & 0.35 & 16.23 \\
643 & 17.41 & 4.4 & 0.31 & 21.47 \\
644 & 17.31 & 3.9 & 0.28 & 18.46 \\
645 & 17.42 & 3.8 & 0.27 & 20.72 \\
646 & 17.23 & 4.0 & 0.28 & 17.68 \\
647 & 17.46 & 5.2 & 0.37 & 20.02 \\
648 & 17.15 & 6.2 & 0.44 & 17.95 \\
649 & 17.05 & 7.3 & 0.52 & 16.68 \\
650 & 17.15 & 7.4 & 0.53 & 11.86 \\
708 & 17.21 & 4.2 & 0.30 & 20.12 \\
709 & 17.46 & 5.7 & 0.41 & 20.07 \\
710 & 17.62 & 4.8 & 0.34 & 22.91 \\
711 & 17.11 & 3.5 & 0.25 & 19.74 \\
715 & 17.35 & 4.1 & 0.29 & 21.52 \\
716 & 16.78 & 3.2 & 0.23 & 18.71 \\
717 & 17.14 & 2.5 & 0.17 & 17.64 \\
718 & 17.35 & 3.4 & 0.24 & 20.39 \\
719 & 17.25 & 2.5 & 0.18 & 20.20 \\
720 & 17.30 & 3.7 & 0.26 & 19.85 \\
721 & 17.14 & 3.6 & 0.25 & 19.39 \\
722 & 16.86 & 5.8 & 0.41 & 18.95 \\
\hline
\end{tabular}
\end{table}

\begin{figure}[tb]
\epsfxsize=\hsize
\epsfclipon
\epsfbox[72 86 341 257]{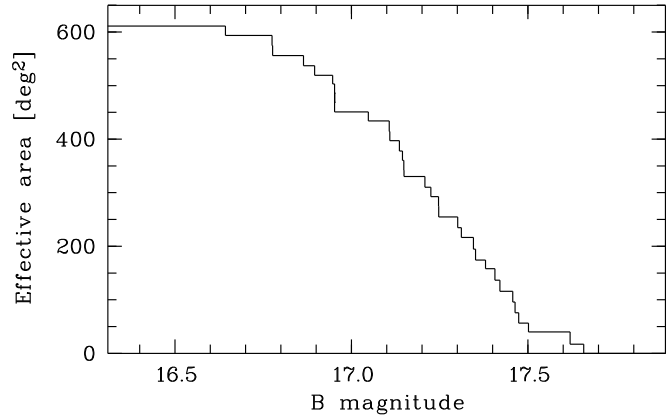} 
\caption[]{Effective area of the surveyed region
as a function of $B$ magnitude (without correction
for Galactic extinction).}
\label{fig:effarea}
\end{figure}

\begin{table}[tb]
\caption[]{The flux-limited sample of QSOs with
$z > 0.07$.
The entries in column $B$ marked by colons are photographic,
all other magnitudes are CCD measurements corrected to the
zero-points of the photographic plates.
Column $B_0$ lists the extinction-corrected magnitudes.}
\label{tab:qsos}
\begin{tabular}{llllll}
Name           &  $z$  & $B$ & Field & $B_0$ & $M_B$ \\[0.3ex]
\hline
HE 0952$-$1552 & 0.108 & 16.56 & 637 & 16.17 & $-$22.87 \rule{0em}{2.7ex}\\
HE 1006$-$1211 & 0.693 & 16.37 & 709 & 15.96 & $-$27.03 \\
HE 1007$-$1405 & 0.583 & 16.05 & 637 & 15.66 & $-$26.94 \\
HE 1012$-$1637 & 0.433 & 16.22 & 638 & 15.69 & $-$26.25 \\
HE 1015$-$1618 & 0.247 & 15.91:& 638 & 15.38 & $-$25.40 \\
HE 1019$-$1413 & 0.077 & 16.75 & 638 & 16.22 & $-$22.09 \\
PKS 1020$-$103 & 0.197 & 17.39 & 710 & 17.05 & $-$23.26 \\
HE 1021$-$0738 & 1.800 & 17.47 & 710 & 17.12 & $-$28.00 \\
PKS 1022$-$102 & 2.000 & 17.58 & 710 & 17.24 & $-$28.14 \\
HE 1025$-$1915 & 0.323 & 16.91 & 568 & 16.50 & $-$24.84 \\
HE 1029$-$1401 & 0.086 & 14.07 & 638 & 13.54 & $-$25.01 \\
HE 1031$-$1457 & 0.652 & 17.39 & 639 & 17.03 & $-$25.83 \\
HE 1041$-$1447 & 1.569 & 17.26:& 639 & 16.90 & $-$27.91 \\
HE 1043$-$1443 & 0.599 & 17.06 & 639 & 16.70 & $-$25.96 \\
HE 1045$-$2322 & 0.407 & 16.65 & 501 & 16.26 & $-$25.55 \\
PKS 1048$-$090 & 0.345 & 16.51 & 711 & 16.27 & $-$25.21 \\
HE 1104$-$1805 & 2.319 & 16.30 & 570 & 16.00 & $-$29.62 \\
HE 1109$-$1255 & 0.596 & 17.19 & 640 & 16.84 & $-$25.80 \\
HE 1110$-$1910 & 0.111 & 16.77 & 570 & 16.47 & $-$22.63 \\
HE 1115$-$1735 & 0.217 & 16.25 & 570 & 15.95 & $-$24.56 \\
HE 1120$-$2713 & 0.389 & 16.94 & 503 & 16.60 & $-$25.12 \\
HE 1159$-$1338 & 0.506 & 16.68 & 643 & 16.36 & $-$25.91 \\
HE 1200$-$1234 & 0.553 & 17.22 & 643 & 16.91 & $-$25.56 \\
HE 1201$-$2409 & 0.137 & 16.74 & 505 & 16.21 & $-$23.33 \\
HE 1211$-$1322 & 1.125 & 16.15 & 644 & 15.87 & $-$28.25 \\
HE 1223$-$1543 & 1.735 & 17.14 & 644 & 16.86 & $-$28.17 \\
HE 1228$-$1637 & 0.102 & 16.78 & 644 & 16.50 & $-$22.43 \\
HE 1233$-$2313 & 0.238 & 17.06 & 506 & 16.52 & $-$24.18 \\
HE 1237$-$2252 & 0.096 & 17.29 & 506 & 16.75 & $-$22.05 \\
HE 1239$-$2426 & 0.082 & 16.73 & 507 & 16.22 & $-$22.25 \\
HE 1254$-$0934 & 0.139 & 15.72 & 718 & 15.48 & $-$24.11 \\
HE 1255$-$2231 & 0.492 & 17.02 & 507 & 16.51 & $-$25.71 \\
HE 1258$-$0823 & 1.153 & 16.44 & 718 & 16.20 & $-$27.97 \\
HE 1258$-$1627 & 1.709 & 17.22 & 646 & 16.94 & $-$28.06 \\
PKS 1302$-$102 & 0.278 & 15.15 & 718 & 14.91 & $-$26.12 \\
HE 1304$-$1157 & 0.294 & 17.14 & 718 & 16.90 & $-$24.25 \\
HE 1312$-$1200 & 0.327 & 16.07 & 719 & 15.89 & $-$25.48 \\
HE 1315$-$1028 & 0.099 & 16.75 & 719 & 16.57 & $-$22.28 \\
Q 1316$-$0734  & 0.538 & 16.70 & 719 & 16.52 & $-$25.89 \\
HE 1335$-$0847 & 0.080 & 16.80 & 720 & 16.54 & $-$21.86 \\
HE 1341$-$1020 & 2.134 & 17.28 & 720 & 17.02 & $-$28.48 \\
HE 1345$-$0756 & 0.777 & 17.13 & 720 & 16.87 & $-$26.40 \\
HE 1358$-$1157 & 0.408 & 16.64:& 721 & 16.39 & $-$25.43 \\
HE 1403$-$1137 & 0.589 & 16.86 & 721 & 16.61 & $-$26.01 \\
HE 1405$-$1545 & 0.194 & 16.51 & 649 & 15.99 & $-$24.29 \\
HE 1405$-$1722 & 0.661 & 15.85 & 649 & 15.33 & $-$27.55 \\
PG 1416$-$1256 & 0.129 & 16.88 & 650 & 16.35 & $-$23.07 \\
HE 1419$-$1048 & 0.265 & 16.77 & 722 & 16.36 & $-$24.58 \\[0.3ex]
\hline
\end{tabular}
\end{table}

\section{The quasar sample}

\subsection{Survey area}

The investigated area was defined by 33 ESO Schmidt fields 
located in the region 
$-28\degr <\delta < -7\degr$ and $9^{\subs{h}}\,40^{\subs{m}}<\alpha <
14^{\subs{h}}\,40^{\subs{m}}$, thus all in the North Galactic
hemisphere. A list of the survey fields and their properties 
is given in Table~\ref{tab:fields}.

In each field, one direct plate from the ESO (B) atlas, and
one HES spectral (objective prism) plate were available.
The formal total area subtended by 33 ESO plates is 
$\sim 25\,\mathrm{deg}^2\times 33 = 825\,\mathrm{deg}^2$.
Losses of usable area occurred because
(1) plates overlapped in adjacent fields, 
especially if not precisely centred; 
(2) occasionally direct and spectral plates were 
positionally mismatched, and only the common area could
be used;
(3) most important, overlapping spectra (and to a lesser
degree, direct images) rendered a certain percentage of
spectra unprocessable.
These losses have been quantified and 
incorporated into `effective areas' $\Omega_{\subs{eff}}$ 
for each field, given in Table~\ref{tab:fields}.
Because of the field-to-field variations in limiting magnitude, 
the total effective survey area is a
function of apparent magnitude, as plotted in Fig.~\ref{fig:effarea}. 
The maximum area is 611\,deg$^2$ for $B<16.64$ and decreases
gradually to zero for sources fainter than $B=17.66$. 

While many quasar surveys were carried out at very high Galactic
latitudes where foreground extinction corrections are small
(although systematically non-zero), this is not the case
for our present set of fields. We have used the data from
Stark et al.\ (\cite{stark:92}) to obtain the neutral hydrogen
column density in the centre of each field, and converted this
into a $B$ band extinction using the formula $A_B = 4.2 \times N_H /
59$ where $N_H$ is given in units of $10^{20}\,\mathrm{cm}^{-2}$
(cf.\ Spitzer \cite{spitzer:78});
Table~\ref{tab:fields} lists $N_H$ and $A_B$ for each field.
The average extinction is 0.36\,mag.

\subsection{Sample selection}

The limiting brightness of QSO candidates during the automated
selection procedure was initially defined by a minimum S/N ratio of
$\sim 3$ in the digital spectra. We later found, as expected
(cf.\ Paper~I), that close to the selection limit the object lists 
became incomplete. Several experiments showed that 
systematic incompleteness occurred only for the faintest
spectra, and that the effect disappeared for a limiting S/N $\ge$ 5.
We translated this S/N cutoff
into a magnitude limit on the direct plate, using the calibration
of photographic magnitudes described below. For the 33 fields, the 
values for $B_{\subs{lim}}$ range between 17.66 and 16.64,
depending largely on the quality of the spectral plates.

Altogether, 115 QSOs and Seyfert~1 galaxies were detected 
by the HES selection criteria in these fields 
(primarily different measures of UV excess; cf.\ Paper~I)
that were either confirmed by our own follow-up spectroscopy 
(Paper~II), or previously catalogued as QSO or Seyfert~1 
in the compilation of V\'{e}ron-Cetty \& V\'{e}ron (\cite{veron:93}).
Note that in this paper we deliberately ignore the 
traditional subdivision Seyferts vs.\ `real' QSOs based on some
arbitrary luminosity threshold. A lower redshift limit of 0.07
was applied to avoid possible incompleteness due to host galaxy
contamination (but see the $z<0.07$ Seyfert sample below). 
Several of the lowest redshift objects would nevertheless 
be called Seyferts by many others.

There remained 48 QSOs with $z>0.07$ located within the 
effective survey area and
above the completeness limiting magnitude. These form the
flux-limited sample listed in
Table~\ref{tab:qsos}. A comparison with the 
V\'{e}ron-Cetty \& V\'{e}ron (\cite{veron:93}) catalogue
revealed no additional QSOs that should have been included
in the flux-limited sample. This test for `completeness'
is, however, not very strong, as few other surveys have covered 
this region of the sky, and $\sim 90$\,\% of the sample objects
are new discoveries made by the HES.

\begin{table*}[tb]
\caption[]{Seyfert~1 galaxies with
$z < 0.07$. The CCD magnitudes
are for the small aperture (see text).
Absolute magnitudes are given
as observed ($M_{B,\subs{obs}}$), 
and with host galaxy contribution $M_{\subs{gal}}$ 
subtracted ($M_{\subs{nuc}}$).
}
\label{tab:seyferts}
\begin{tabular}{llllllllll}
Name &  $z$  & $B$ & Field & $B_0$ & $z_{\subs{max}}$ &
      $V_{\subs{e}}/V_{\subs{a}}$ &
      $M_{B,\subs{obs}}$ & $M_{\subs{gal}}$ & $M_{\subs{nuc}}$ \\[0.3ex]
\hline
HE 1043$-$1346 & 0.0669 & 16.83 & 639 & 16.47 & 0.0692 & 0.902 & $-$21.55 & $-$21.06 & $-$20.44 \rule{0em}{2.7ex}\\
HE 1248$-$1357 & 0.0144 & 15.28 & 645 & 15.01 & 0.0152 & 0.855 & $-$19.67 & $-$19.39 & $-$18.06 \\
IRAS 1249$-$13 & 0.0129 & 14.91 & 645 & 14.64 & 0.0254 & 0.132 & $-$19.80 & $-$19.10 & $-$18.99 \\
R 12.01        & 0.0463 & 15.50 & 646 & 15.21 & 0.0700 & 0.289 & $-$22.00 & $-$20.90 & $-$21.52 \\
PG 1310$-$1051 & 0.0337 & 15.64 & 719 & 15.46 & 0.0544 & 0.238 & $-$21.07 & $-$19.54 & $-$20.77 \\
HE 1330$-$1013 & 0.0221 & 15.82 & 719 & 15.65 & 0.0293 & 0.428 & $-$19.96 & $-$18.96 & $-$19.41 \\
R 14.01        & 0.0408 & 14.68 & 648 & 14.24 & 0.0700 & 0.196 & $-$22.70 & $-$20.72 & $-$22.51 \\[0.3ex]
\hline
\end{tabular}
\end{table*}

\subsection{Photometry}

Obtaining consistent and unbiased photometry of low-redshift quasars 
with detectable host galaxies is not trivial. 
During the survey phase, normally only photographic data, in our case from the
digitised direct $B$ plates, are available.
These plates usually reach much fainter magnitudes than
objective-prism plates, but they also saturate earlier,
especially if not the original plates but glass copies
are used (as was the case for several of our fields).
The conventionally employed isophotal or large aperture magnitudes,
although well suited for point sources,
tend to give much larger total luminosities for
QSOs with detectable hosts than desired for
an investigation of the {\em nuclear\/} luminosity function.
Additional complications can result from non-linear
filtering effects inherent in the digitisation procedure.

For the present sample we have developed a two-step procedure 
to obtain unbiased and survey-consistent magnitudes.
The photographic plates were calibrated by CCD sequences
obtained with the ESO 90\,cm telescope in 1993 and 1994.
These sequences will be published in a separate paper.
The sample selection was based on
simulated diaphragm photometry with an aperture 
of the size of the seeing disk.
For point sources, an accurate calibration relation 
with very little intrinsic scatter, $\sim 0.15$ mag,
could be derived from the CCD sequences 
(which contained only stars). 
These small-aperture magnitudes
correlated also well with the corresponding S/N ratios of
the digital spectra, important for an accurate definition
of limiting magnitudes.

In a second step, we obtained CCD photometry 
for most of the sample QSOs.
Exposures in the $B$ band were available from the 
mandatory acquisition images 
at the ESO 3.6\,m and 2.2\,m telescopes (cf.\ Paper~II).
After standard reduction and colour term corrections,
the photometric zero points of these images were shifted
to the system of the corresponding photographic plate
by using surrounding unsaturated field stars as references.
The resulting corrected CCD magnitudes,
again measured in an aperture of diameter
approximately equal to that of the seeing disk,
should be much more accurate than the photographic data,
at least for extended  objects,
while magnitudes of point sources should be more or less unchanged.
We tested this assertion by comparing photographic
and zero-point corrected CCD photometry of the 
$z > 0.3$ QSOs (which should be true point sources)
and found a mean difference of only 0.01\,mag.
For the further analysis we used the zero-point corrected
$B$ band CCD magnitudes. For three
objects no CCD data were available, and for those we substituted the
photographic measurements (cf.\ Table~\ref{tab:qsos}).

\subsection{The additional Seyfert sample}

To obtain further insight into the continuity of properties
between high-luminosity QSOs and low-luminosity Seyfert nuclei,
we conducted a dedicated subsurvey for Seyfert galaxies, targeting
both type~1 and type~2 objects. The criteria were different from those
used in the main survey, adapted to discriminate between active
and inactive galaxies rather than between stars and QSOs. 
More details about this subsurvey will be given in a later paper
(K\"ohler et al., in prep.)
where we shall also present our results
concerning Seyfert~2 galaxies, which we do not consider here.

We found 7 Seyfert~1 galaxies with $0.01<z<0.07$,
listed in Table~\ref{tab:seyferts}. The redshifts were
remeasured in new spectra obtained with the ESO 1.52~m telescope,
and corrected for heliocentric motion.
Not all the fields from Table~\ref{tab:fields} 
were involved in this subsurvey, 
and the effective area was only 477 deg$^2$ (but see
Sect.\ \ref{sec:sy1lf} for the computation of space densities).

For these sources, proper photometry is at least as important
as for QSOs.
The photometric systems of many earlier investigations 
of the luminosity function of Seyferts suffer 
from two severe drawbacks: (i) They were often based on heterogeneous
collections of published (aperture) photometry, 
generally inconsistent with the magnitudes used for defining the 
survey flux limits.
(ii) The contributions of host galaxies were either not corrected for
at all, or some assumptions about universal intrinsic properties
of hosts and nuclei had to be made, such as colours
(Sandage \cite{sandage:73}) or host luminosities
(Cheng et al.\ \cite{cheng:85}).

On the assumption that Seyfert~1 galaxies are the low-luminosity equivalent
of quasars, the quantity of interest is the {\em nuclear\/}
brightness, with the host galaxy contribution removed. 
Although our small-aperture magnitudes were already much less affected
by galaxy contamination than large-aperture or total brightness
measurements,
a major contribution could still be expected especially for the lowest
nuclear luminosities. We therefore estimated 
these corrections individually, based on
$B$ and $V$ images of the seven sample objects 
taken with the ESO 90\,cm telescope.
We first constructed empirical two-dimensional 
point-spread functions from nearby stars. These were then
subtracted from the Seyfert images, scaled by a factor
chosen such that the residual galaxy surface brightness
distribution did not decrease inwards. The resulting
host galaxy brightness was finally integrated over the aperture 
and subtracted from the total aperture flux. 
As can be seen in Table~\ref{tab:seyferts}, the corrections were not
negligible, and -- expectedly -- largest for the 
lowest redshift sources.

\section{The surface density of bright QSOs}

\begin{figure}[tb]
\epsfxsize=\hsize
\epsfclipon
\epsfbox[68 86 341 257]{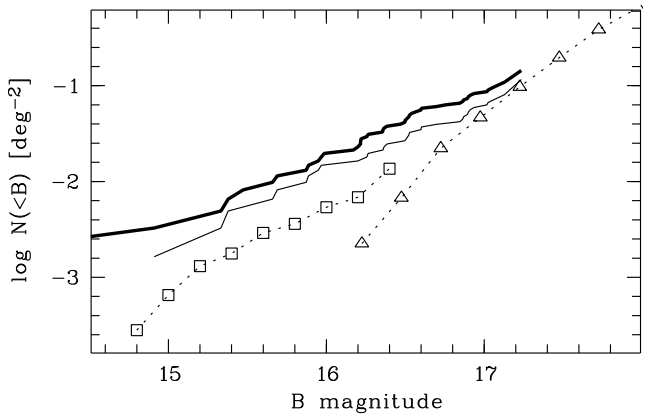} 
\caption[]{Cumulative surface densities of bright QSOs with $z<2.2$. 
Thick line: this work, for $z>0.07$, thin line: the same for $z>0.2$. 
Open squares show the relation from Schmidt \& Green 
(\cite{schmidt:83}), modified for $z>0.07$ (see text).
Triangles give the LBQS (Hewett et al.\ \cite{hewett:95}) 
relation, valid for $z>0.2$.}
\label{fig:surfdens}
\end{figure}

An important diagnostic tool for QSO samples is
the empirical relation between source fluxes and the
number of sources per unit solid angle. Once the field
properties and the sample photometry are complete, the
cumulative surface density of QSOs brighter than magnitude $B$,
$N(<B)$, can be easily computed from summing over 
$1/\Omega_{\subs{eff}}$ for all relevant sample objects.
From our QSO sample with $z<2.2$ (47 objects) we 
have derived the $N(<B)$ relation shown as thick line in
Fig.~\ref{fig:surfdens}. Note that the abscissa values
in this diagram are extinction-corrected magnitudes.
The bright end of the curve is dominated
by the extremely luminous source HE~1029$-$1401, but the main part
can be well approximated by a simple power law of slope 
$\beta = 0.67\pm 0.04$ ($\log N = \beta B + \mathrm{const}$).
If only the 33 sources with $z>0.2$ are considered, excluding thereby
all QSO/Seyfert borderline cases (thin line in Fig.~\ref{fig:surfdens}), 
the normalisation drops slightly but the slope is essentially
unchanged with $\beta = 0.69\pm 0.05$.
For those readers who prefer differential surface densities, 
we have computed the figures in half magnitude bins between $B=14.75$
and $B=17.25$ (Table~\ref{tab:surfdens}). 
The counts were multiplied by a factor of 2 to formally express 
number counts per unit magnitude intervals.

Comparison relations from other surveys 
are numerous for fainter magnitudes, but rare for 
$B\la 16.5$. The Palomar Bright Quasar Survey 
(BQS; Schmidt \& Green \cite{schmidt:83}, hereafter quoted as SG83)
provides the only available optically selected QSO sample 
with a well-defined flux limit subtending over a similar region 
in the Hubble diagram.
SG83 listed differential surface densities
computed from the full set of 114 QSOs with $z<2.2$ found
over an area of 10\,714 deg$^2$. To make their numbers comparable
to ours, we recomputed the BQS surface densities by removing
all $z<0.07$ entries from their sample, and converted the result into a 
cumulative relation. This relation is shown in
Fig.~\ref{fig:surfdens} by the open squares. A large discrepancy at
all magnitudes is apparent. At $B=16$, the surface density
found in the HES is high by a factor of 3.6 compared to the BQS,
certainly more than can be permitted by statistical fluctuations.

\begin{table}[tb]
\caption[]{Differential surface density $A(B)$ of 33 HES QSOs
with redshifts $0.2 < z < 2.2$, expressed
as number of QSOs per deg$^2$ per unit magnitude interval.
$n$ gives the actually observed number per half magnitude
bin ($B\pm 0.25$).}
\label{tab:surfdens}
\begin{tabular}{llll}
$B$  &  $n$ & $A(B)$ & $\sigma_A$ \\[0.3ex]
\hline
15.0  &  1   & 0.003 & 0.003 \rule{0em}{2.7ex} \\
15.5  &  4   & 0.013 & 0.007 \\
16.0  &  5   & 0.016 & 0.008 \\
16.5  & 12   & 0.046 & 0.014 \\
17.0  & 11   & 0.15  & 0.08  \\[0.3ex]
\hline
\end{tabular}
\end{table}

A similar result was already reported by Goldschmidt et al.\ 
(\cite{goldschmidt:92}) from an analysis of the Edinburgh survey, 
who found 5 QSOs above the BQS flux limit
in an area of 330\,deg$^2$ (non-overlapping with the present HES area)
while the BQS contained only one.
Their estimate of $N(B<16.5)=0.024$\,deg$^{-2}$ for $0.3<z<2.2$ 
is completely consistent with our corresponding value of 
$0.021\pm 0.05$. Furthermore, Goldschmidt et al.\ found a
zero-point offset of 0.28\,mag between the 
BQS photographic magnitudes and theirs,
in the sense that the Palomar measurements were systematically 
{\em too bright\/}, thus increasing rather than relaxing the discrepancy. 

It should be noted that neither SG83 nor Goldschmidt et al.\
corrected their magnitudes for Galactic extinction, whereas we have
done so.
However, our survey area is {\em predominantly\/} located in
regions of rather high $A_B$, while SG83 estimated that the extinction
for most BQS quasars were `not much more than 0.1\,mag';
this even might be, at least partly,
compensated by the above mentioned zero-point offset. 
Because of these uncertainties we have not attempted to
correct the BQS results for extinction. One can estimate globally
that an average $A_B$(BQS) of 0.61\,mag would be necessary
to bring BQS and HES surface densities to a match -- 
a totally unrealistic value.

At fainter magnitudes the agreement of HES number counts 
with those of other surveys is excellent.
As an example, we have plotted the values from the LBQS
(Hewett et al.\ \cite{hewett:95}) into Fig.~\ref{fig:surfdens},
with the abscissa shifted by +0.1\,mag to transform their $B_J$
photometric system approximately into $B$ magnitudes. While
the counts in the brightest bins are again lower than ours,
this can be understood as a result of poor statistics in the 
LBQS for $B\la 16.5$, possibly arising from saturation effects.  
Around $B\simeq 17$, HES and LBQS surface densities join smoothly,
without any detectable significant offset. Similarly good
agreement is reached with other surveys, cf.\ the recent compilation
by Cristiani et al.\ \cite{cristiani:95}.
We are thus confident that the photometric
scale used in the present investigation is essentially unbiased 
and adequate for luminosity function work.

\begin{figure}[tb]
\epsfxsize=\hsize
\epsfclipon
\epsfbox[68 86 341 257]{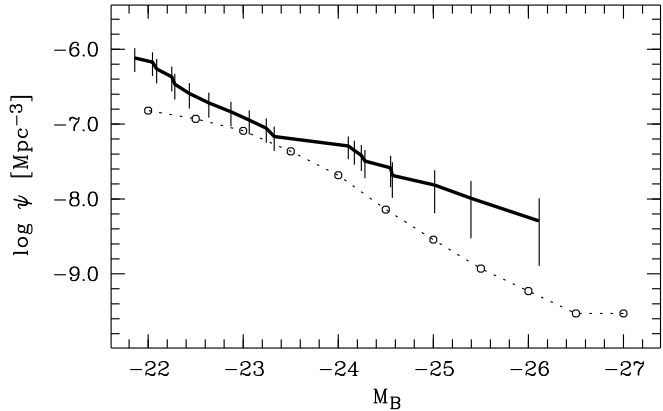} 
\caption[]{Cumulative luminosity function of QSOs 
with $0.07 < z < 0.3$ (thick line).
For comparison: Cumulative luminosity function of BQS quasars 
with $0.07 < z < 0.3$, constructed
as described in the text
(dotted line and small symbols).}
\label{fig:lqlf}
\end{figure}

\begin{figure}[tb]
\epsfxsize=\hsize
\epsfclipon
\epsfbox[68 86 341 257]{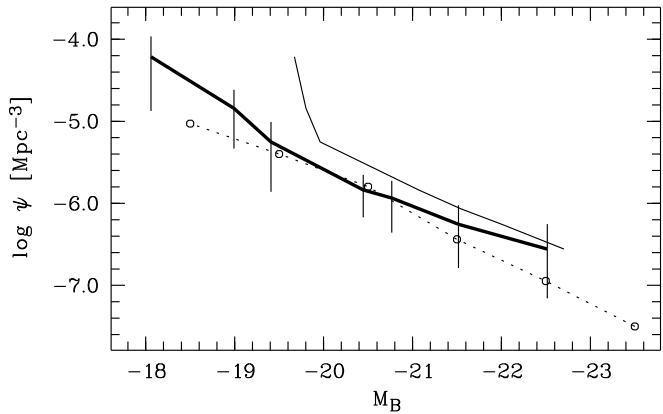} 
\caption[]{Cumulative luminosity function of the 
additional Seyfert~1 sample ($z<0.07$). Thick line:
LF for nuclear magnitudes corrected for 
the host galaxy contributions; thin line:
uncorrected LF. For comparison:
Seyfert~1 nuclear luminosity function of Cheng et al.\ 
(\cite{cheng:85}; dotted line and small symbols)}
\label{fig:sy1lf}
\end{figure}

\section{Luminosity functions}

\subsection{The luminosity function of low-redshift QSOs}

For all objects in Table~\ref{tab:qsos}, we first estimated 
absolute magnitudes for the rest-frame $B$ band by computing their
luminosity distances in an expanding Friedmann universe
with the formula of Terrell (\cite{terrell:77}), and applying
$K$ corrections taken from Cristiani \& Vio (\cite{cristiani:90}).
As low-redshift subsample we selected the 20 QSOs with $z<0.3$.
Space densities were derived using the generalised $1/V_{\subs{max}}$ 
estimator (Felten \cite{felten:76}; Avni \& Bahcall \cite{avni:80}), 
thus incorporating the full information about field-dependent 
flux limits, including fields with zero detections
(the `coherent' analysis in the terminology of Avni \& Bahcall). 

The resulting cumulative luminosity function is shown 
in Fig.~\ref{fig:lqlf}. 
Locations of individual objects are indicated by the error bars.
We have chosen the cumulative representation $\psi(<M_B)$
as it allows one to avoid binning (particularly problematic for
smaller samples) and to make the contribution of each 
object in the sample apparent.
The errors were estimated from Poisson statistics in the sample, 
combined with the uncertainties in volume 
determination resulting from the photometric errors. 

Restricting the HES luminosity function to $z<0.2$
(14 objects) results in slightly larger error bars, 
but the space densities change only insignificantly, 
by less than 0.05\,dex, and, overplotted in Fig.~\ref{fig:lqlf},
could not even be distinguished from the $z<0.3$ relation.
We thus feel justified to neglect differential evolution within 
our redshift shell; we shall further investigate and 
qualify this assumption below.

Up to now, the BQS provides the only comparison sample
in the region $M_B < -23$ and $z<0.3$, containing  45 objects 
that fulfill also the condition $z>0.07$. Surprisingly,
there is yet -- to our knowledge -- no {\em direct\/} determination 
of the local QLF based on BQS objects alone.
We have therefore computed a QLF from this sample with the
same methods used for our own data, taking magnitudes,
survey limits, and areas from SG83.
We show the results of these computations
as a cumulative distribution in Fig.~\ref{fig:lqlf}.
(Since all other literature data are available only in 
binned form, we have also binned the BQS LF in $M_B$,
to achieve a more uniform presentation.)
As mentioned above, Galactic extinction is neglected for
these data; at any rate, the effect should be very small.

The space densities found from the HES data are much
higher than those derived from the BQS, with the discrepancy 
increasing with luminosity. For the most luminous sources, 
we find that there are almost an order of magnitude more
low-redshift quasars per unit volume than discovered by the BQS.
Possible reasons for this discrepancy are discussed in
Sect.\ \ref{sec:discussion}.

\subsection{The space density of Seyfert~1 nuclei \label{sec:sy1lf}}

To extend the local luminosity function towards fainter levels
of nuclear activity, we have used the additional Seyfert~1 sample
with $z<0.07$. The absolute magnitudes were computed
in the approximation of Euclidean static geometry and neglecting
$K$ corrections. To obtain reliable and unbiased 
estimates for the accessible volumes, 
the maximum redshift $z_{\subs{max,$i$}}$ at which
object $i$ would still have been included in the survey
had to be determined for each object. Because of the superposition
of a host galaxy with constant intrinsic scale length 
and a point-like nucleus,
the small-aperture magnitudes depended on redshift in a complicated
way, and $z_{\subs{max,$i$}}$ was therefore not only a function of $B$ and
$B_{\subs{lim}}$ but also of intrinsic properties. We used
the available CCD images to simulate the mixing
of AGN\,/\,host contributions into the aperture
as a function of redshift $z$, resulting in an array of 
$z_{\subs{max,$k$}}$ values, one for each field $k$.
For simplicity, Table~\ref{tab:seyferts} supplies 
`effective' values $z_{\subs{max}}$
that permit a direct conversion into space densities.
A $V/V_{\subs{max}}$ test (Schmidt \cite{schmidt:68}), 
implemented in the generalisation
proposed by Avni \& Bahcall (\cite{avni:80}), gives a mean value of 
$V/V_{\subs{max}} = 0.43\pm 0.11$
(individual values are also listed in Table~\ref{tab:seyferts}).
Compared with the expectation value of 0.5 for a complete,
non-evolving, homogeneously distributed sample,
there is no evidence for major incompleteness.

The resulting luminosity function is displayed in
Fig.~\ref{fig:sy1lf}.
It can be seen that a host galaxy correction is important
particularly for the intrinsically faintest objects in the sample:
Without correction, the luminosity function shows a steep upturn 
at $M_B \simeq -20$; for these objects, even the small-aperture 
magnitude is dominated by the host rather than by the AGN.

Although the sample is not large, the derived space
densities are fully consistent with previous estimates
(Cheng et al.\ \cite{cheng:85}; Huchra \& Burg \cite{huchra:92}). 
However, none of these previous investigations have payed 
as much attention to an appropriate host galaxy correction, 
neither for source luminosities nor for the survey volume,
as we have. In particular, the analysis of the CfA sample 
by Huchra \& Burg was based on uncorrected Zwicky magnitudes
(essentially measuring the {\em total\/} galaxy brightness),
making a straightforward comparison impossible. 
More similar to our approach was the work of Cheng et al.\ (\cite{cheng:85})
with Markarian galaxies, but their sample, although much larger than ours,
had to be manipulated with substantial incompleteness corrections,
in addition to the afore mentioned problems of a heterogeneous
photometric data base.

\begin{figure}[tb]
\epsfxsize=\hsize
\epsfclipon
\epsfbox[68 86 341 257]{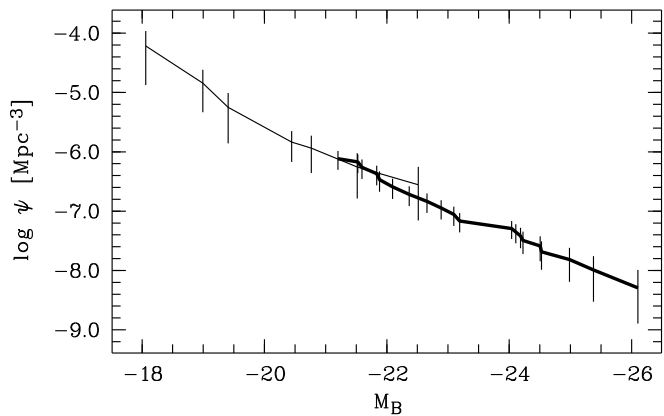} 
\caption[]{Combined local luminosity function of
quasars and Seyfert~1 nuclei with $z<0.3$, 
based on magnitudes corrected for the host galaxies. }
\label{fig:syqlf}
\end{figure}

\begin{table}[tb]
\caption[]{Binned differential local luminosity function, 
computed from 27 QSOs and Seyferts with $z < 0.3$, 
in Mpc$^{-3}$ per unit absolute magnitude interval
centred on the value given by $M_B$.
Column $n$ gives the number of objects per bin.}
\label{tab:lqlf}
\begin{tabular}{llll}
$M_B$  &  $n$ & $\phi(M_B)$ & $\log\phi$ \\[0.3ex]
\hline
-18.5  &  2   & $(5.5 \pm 4.7 ) \times 10^{-5}$ & $-4.3$ \rule{0em}{2.7ex}  \\
-19.5  &  1   & $(4.1 \pm 4.1 ) \times 10^{-6}$ & $-5.4$ \\
-20.5  &  2   & $(9.0 \pm 6.7 ) \times 10^{-7}$ & $-6.0$ \\
-21.5  &  2   & $(1.1 \pm 0.9 ) \times 10^{-7}$ & $-7.0$ \\
-22.5  &  8   & $(4.5 \pm 1.8 ) \times 10^{-7}$ & $-6.4$ \\
-23.5  &  3   & $(6.2 \pm 3.6 ) \times 10^{-8}$ & $-7.2$ \\
-24.5  &  6   & $(3.6 \pm 1.5 ) \times 10^{-8}$ & $-7.5$ \\
-25.5  &  2   & $(1.0 \pm 0.7 ) \times 10^{-8}$ & $-8.0$ \\
-26.5  &  1   & $(5.0 \pm 5.0 ) \times 10^{-9}$ & $-8.3$  \\[0.3ex]
\hline
\end{tabular}
\end{table}

\subsection{The combined local luminosity function}

The local universe is the only domain where it is possible 
to construct the entire QLF, including low-luminosity tail, 
directly from one single survey.
The local luminosity function of QSOs and Seyfert~1 nuclei,
obtained by combining the two independently derived luminosity
functions presented above, is valid between absolute magnitudes 
$-18$ and $-26$, thus spanning three decades in luminosity.

Overlaying the cumulative relations of Figs.\ \ref{fig:lqlf}
and \ref{fig:sy1lf}, the match in the common luminosity region 
is remarkably good (cf.\ Fig.~\ref{fig:syqlf}). To arrive
at a consistent system of {\em nuclear\/} luminosities, 
the magnitudes in the QSO sample were corrected by 
subtracting a template host galaxy of $M_B = -21$. 
Since the magnitudes had already been measured through a small 
aperture, this host subtraction was a minor correction for
all objects in the QSO sample, 
and the results are not sensitive to the assumed host luminosity.

The combined local QLF can be fitted well by a simple linear
relation between $M_B$ and $\log \psi (M)$, corresponding to 
a single power-law for the {\em differential\/} QLF 
($\phi (L) dL \propto L^\alpha$). 
A slope parameter $\alpha = -2.17\pm 0.06$
provides a statistically acceptable fit 
over the full range of 8 magnitudes. 
Fitting only the QSO sample ($z>0.07$) does not alter the slope 
significantly ($\alpha = -2.10\pm 0.08$);
fitting only the $z<0.07$ Seyferts yields $\alpha = -2.38\pm 0.23$,
again consistent with a constant slope of $\alpha \simeq -2.2$.
The slope of the BQS LF is much steeper:
Apart from flattening at both high and low luminosities, 
the data demand $\alpha \la -3$ over the most of the range.

From the combined dataset, we have also produced a binned differential
luminosity function $\phi (M)$, 
tabulated in steps of one in absolute magnitude
(Table~\ref{tab:lqlf}). As was to be expected, this representation
is quite sensitive to the actual choice of binning intervals,
most obvious in the succession of a `high' and a `low' bin
at $M_B = -22.5$ and $-21.5$, respectively. There is, however,
full consistency of the results obtained; in particular,
the QLF slope parameter $\alpha \simeq 2.2$
is reproduced within the expected uncertainties.

\begin{figure}[tb]
\epsfxsize=\hsize
\epsfclipon
\epsfbox[68 86 341 257]{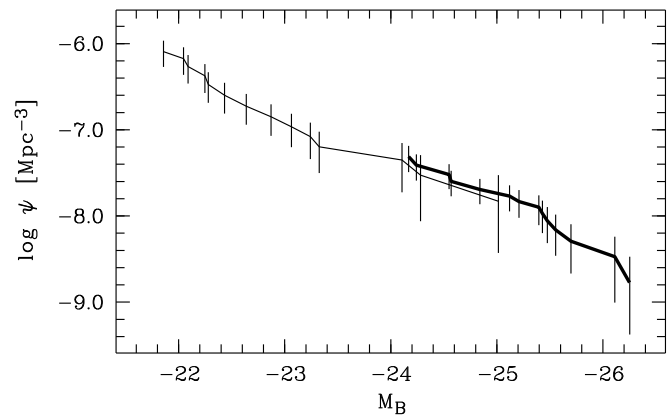} 
\caption[]{The bright end of the quasar luminosity function up to $z=0.5$, 
in comparison with the local QLF. Shown are the cumulative
relations in the redshift shells $z<0.2$ (thin line),
and $0.2<z<0.5$ (thick line).}
\label{fig:evolqlf}
\end{figure}

\subsection{The quasar luminosity function at higher redshifts}

The sample of Table~\ref{tab:qsos} contains eight QSOs with 
redshifts $0.3<z<0.5$ that we used to further constrain the potential
error introduced by neglecting differential evolution.
Binning the full $z<0.5$ sample into two redshift regimes $0.07<z<0.2$ and
$z>0.2$, respectively, there are 14 objects in each shell.
The luminosity functions are compared in Fig.\ \ref{fig:evolqlf}:
For $-25\la M_B \la -24$, the local and the adjacent $0.2<z<0.5$ 
shell have essentially identical space densities, while
for $M_B<-25$, the QLF in $z>0.2$ seems to be merely a continuation
of the local QLF to higher luminosities, {\em with the same slope
and normalisation\/}. Dividing the sample at $z=0.3$ instead of
0.2 does not affect the results.

Outside the local universe, a single survey can yield no more 
than a small segment of the QLF, and only the combination 
of many different surveys can provide similar
coverage of source luminosities as in the local QLF.
The HES has the potential to provide new samples
covering the brightest parts of the known QSO population,
and it will be a promising task for the future to combine our
results with those of fainter surveys. Such an analysis is,
however, not the scope of the present paper. We only remark that
even at redshifts higher than $z=0.5$, the QLF pieces sampled
by the HES do not show significant deviations from an extrapolation
of the local relation. A more quantitative discussion of this
issue is deferred to a later paper.

\section{Discussion \label{sec:discussion}}

\subsection{Incompleteness of the BQS}

Our results imply that the Palomar Bright Quasar Survey
is not only incomplete, but also heavily biased.
The general incompleteness may be estimated
from the surface densities to be around a factor of 2--3.
This effect has been noted by others before
(Wampler \& Ponz \cite{wampler:85}; 
Goldschmidt et al.\ \cite{goldschmidt:92}),
and we have at present no easy explanation for it.
Some mild adjustment of the BQS data to higher values 
of space densities and luminosities may be required 
because of neglected Galactic extinction, 
but this cannot account for the observed discrepany.
One possibility is that the rather large photometric errors
in the BQS, reflected in the $(U-B)$ colours, 
scattered many QSOs out of the UV excess domain.

There seems to be an additional deficit of luminous
low-redshift QSOs, by another factor of $\ga 2$ for $M_B<-24$.
This discrepancy is apparent already from simply counting 
the number of $z<0.3$ QSOs more luminous than $M_B = -24.5$: 
There are 5 objects in the present HES sample, while
the entire BQS contains only 15 such sources,
in a more than 15 times larger area.
As this is the first detection of such an effect,
one has to carefully consider systematic errors. 
Note, however, that
the HES is the first optical QSO survey since completion of 
the BQS that is sensitive to a similar range of redshifts and
luminosities, and it is therefore not altogether inconceivable
that substantial selection biases have remained unnoticed
over a long time.

The good agreement between the HES $z>0.2$ or $z>0.3$
surface densities with other major surveys, e.g.,
the LBQS (Hewett et al.\ \cite{hewett:95}),
the Edinburgh survey (Goldschmidt et al.\ \cite{goldschmidt:92}), 
or the HBQS (Cristiani et al.\ \cite{cristiani:95}), 
makes it implausible that the HES photometric zeropoint
could be seriously in error.

We also tested our method of zero-point corrected CCD photometry.
All relevant quantities were recomputed based on
the standard approach of using only photographic magnitudes 
measured over large apertures, without any 
systematic change of results, except that the luminosities of 
low-redshift QSOs with bright host galaxies were
often {\em over\/}estimated in the photographic data,
creating a bias towards even higher QLF values.
The CCD approach thus seems justified and appropriate.

The reconstruction of the $0.07<z<0.3$ luminosity function 
for the BQS was based on the same procedures 
as applied to our own data.
We neglected differential evolution within the $z<0.3$
shell, but the {\em observed\/} discrepancy is independent of
any assumed evolution law, and therefore most likely caused
by real incompleteness in the BQS sample. 

The exclusion of objects of non-stellar appearance in the BQS
is an intuitively appealing candidate mechanism for this 
additional incompleteness. It is obvious at first glance
(and has in fact never been disputed) that the BQS sample
is systematically lacking low-luminosity Seyferts; it has
usually been assumed that this selection effect
ceased to be valid for $M_B \la -23$. Our results suggest
that this is not the case. However,
it is then somewhat surprising that the incompleteness should 
be smallest, almost disappearing, in the luminosity 
range around $M_B \simeq -23$, as seems to follow from 
Fig.\ \ref{fig:lqlf}.
In fact, this would then be the {\em only\/} region 
of the entire Hubble diagram where the BQS counts were not 
significantly below those of the HES.

Increased incompleteness for the most luminous QSOs
seems a possible, but rather unlikely explanation.
If, instead, the BQS magnitudes particularly
for the slightly resolved $M_B\simeq -23$ objects 
were {\em systematically too bright\/}, overcompleteness would occur,
scattering objects into the sample from below the flux limit.
We know from our own experience with photometry of
resolved objects on digitised photographic plates that such
effects exist, especially if a large digitisation aperture is 
used (this is why we decided to incorporate the CCD measurements). 
The BQS was based on PDS scans with an aperture of 
$4\farcs 4$ (Green \& Morrill \cite{green:78}), and magnitudes 
were basically isophotal in photographic density.
We {\em suspect\/} that these technical details provide a clue
to the above contradiction, and that incompleteness 
due to the rejection of extended QSO hosts could be
partly counteracted by photometric bias and overcompleteness
for only marginally resolved objects. Further analysis
of this point is certainly required.

If our hypothesis is correct, two important conjectures can be made:
(1) The slope of the QLF for $M_B<-23$ is much flatter
than previous investigators estimated, weakening the evidence 
for a `break' in the Seyfert/QSO transition regime 
(e.g., Marshall \cite{marshall:87}) --
the local QLF presented here showing essentially {\em no\/} break at all.
Note that also in the Seyfert LF of Cheng et al.\ (\cite{cheng:85}),
the presence of such a break is not really demanded by the data.
(2) The BQS sample, up to now the prime source for luminous radio-quiet 
low-redshift QSOs, might suffer from severe selection 
biases with respect to the morphological properties 
of their galaxy hosts.

\subsection{Constraints on quasar evolution}

The local luminosity function derived in this paper is independent
of samples collected at higher redshifts, and therefore well suited
to test and constrain possible evolution laws.
Existing estimates of the QLF at redshift $z\simeq 2$
(e.g., Boyle et al.\ \cite{boyle:91}) show a marked break,
or change of slope, at intermediate luminosities,
that shifts towards lower values with decreasing $z$. 
We have argued above that the local equivalent of this break
is probably an artefact.
The flat slope of the $z\approx 0$ QLF derived in this paper, 
and the absence of clear features, demand,
on the assumption that the high $z$ QLF determinations are correct,
that both shape and slope of the QLF change strongly with redshift.
This clearly is in contradiction to the `standard picture' 
of pure luminosity evolution (PLE), where 
a double power-law luminosity function with invariant logarithmic
shape is merely shifted along the luminosity axis.

However, it should be noted that virtually {\em all\/} 
evolution modelling since 1983 had to rely on the BQS data 
of SG83, to cover an otherwise inaccessible part of the Hubble diagram.
Hence, in these regions, the models merely reproduce the 
incompleteness and possible biases of the BQS contribution.
The problem that QSO surveys were notoriously incomplete at low redshifts 
was of course realised by most researchers, and it has become
customary to apply a lower redshift cutoff of $z_{\subs{min}}\simeq 0.3$
to all samples. The construction of a $z=0$ QLF from such models 
has therefore always
the character of an {\em extrapolation\/} outside the validated range,
and should be treated with appropriate caution. 
Even so, a {\em quantitative\/} comparison between evolutionary models 
and actual measurements of the local Seyfert~1/QLF
is a powerful test which deserves higher attention than 
it has been paid to in the past,
compared to the careful analyses of higher redshift data.

In the construction of the $z<0.3$ local QLF we have not corrected
individual luminosities for statistical evolution.
While it is true that such a neglect could artificially flatten 
the luminosity function, evolution must be fairly strong 
for the effect to be relevant.
We show now that the no-evolution assumption was, in fact, 
the best approximation one could possibly make, over the range
of luminosities covered by the HES. 
In the case of evolution, the LFs of adjacent, non-overlapping
redshift shells should be different. For the simple case of
PLE with the rate suggested, e.g., by Boyle et al.\ (\cite{boyle:91}),
one would expect an abscissa offset of
$\Delta M = -0.8$\,mag between $z<0.2$ and $0.2<z<0.5$.
The necessity to invoke such an offset is not apparent 
from Fig.\ \ref{fig:evolqlf}. While it might be consistent
with our data to allow for a somewhat steeper intrinsic slope of the
QLF, combined with mild evolution, there is certainly 
no {\em evidence\/} for evolution in the data.

These conclusions are supported by results reported recently 
from other surveys. Hewett et al.\ (\cite{hewett:93}),
with a preliminary analysis of the LBQS, 
showed that the QLF flattens considerably at $z<1$;
a similar conclusion was reached by Miller et al.\ 
(\cite{miller:93}) using the Edinburgh survey.
Although both surveys have the usual lower redshift limit 
excluding the local population ($z>0.3$ for the Edinburgh survey,
$z>0.2$ for the LBQS), a prediction for the slope
of the local QLF can be made by extrapolating the trend seen by 
Miller et al.; the result is consistent with
our measured value of $\alpha = -2.2$.
A flatter QLF implies a higher fraction of luminous quasars 
among the total population. Our results indicate therefore,
in agreement with Hewett et al.\ and Miller et al.,
that the evolution of the most luminous quasars must proceed
considerably slower than suggested by the notion
of pure luminosity evolution.

\section{Conclusions and outlook}

We have analysed a new, flux-limited and well-defined 
sample of QSOs \& Seyfert~1 galaxies. Host-galaxy dependent
selection and photometric biases, if not absent, are greatly reduced
in comparison to other optically selected samples.
We find that the derived space density of luminous QSOs
in the local universe is much higher than previous surveys indicated,
and that consequently these objects show a much slower 
cosmological evolution.

With the present sample of 55 QSOs and Seyferts in 33 fields, 
only a small fraction of the HES area is covered. 
We have already acquired,
and partly processed, much additional plate material enlarging the
survey by a significant factor. For example, we now have
several fields in common with the BQS, and it shall soon
be possible to pursue the question of incompleteness and biases 
in the BQS by a direct comparison of objects and photometry.

To understand and model the process of quasar evolution requires
strong constraints from observations. It seems now that we are
farther away from a coherent empirical picture of the QSO population
and its evolution than thought a few years ago. Quasars of different
luminosities evolve differently, and quasars do not necessarily 
evolve with the same rate at different cosmological epochs. 
There are presumably more parameters needed to describe quasar 
evolution properly than in the simple model of 
`pure luminosity evolution'.

The ultimate aim of QSO survey work is the deduction of intrinsic
physical properties from statistical samples. 
Unfortunately the physical
processes in active galaxy nuclei are poorly understood, and the
relations between statistical and physical properties therefore
far from unique. The best study cases for nuclear activity in 
galaxies are nearby Seyferts; as shown above,
the luminosity function of Seyfert~1 nuclei can be smoothly
continued into the classical quasar regime, without a significant
change of slope or break. While this by no means proves that 
Seyfert~1 nuclei are simply scaled-down versions of quasars,
it suggest a continuity of properties rather than
distinct classes. With the completion of the HES, a large
and well-defined sample of highly luminous low-redshift QSOs
will become available, to study the relationship between 
QSOs, their environments, and their evolution in detail.

\begin{acknowledgements}
T.K. acknowledges support from the Deutsche Forschungsgemeinschaft
under grant Re 353/33. Substantial observing time at ESO was
allotted to the project as ESO key programme 02-009-45K.
\end{acknowledgements}

%
%


\begin{thebibliography}{}
%
%
\bibitem[1980]{avni:80}
Avni Y., Bahcall J.N., 1980, ApJ 235, 694
%
\bibitem[1991]{boyle:91}
Boyle B.J., Jones L.R., Shanks T., 1991, MNRAS 251, 482
%
\bibitem[1985]{cheng:85}
Cheng F.-z., Danese L., De Zotti G., Franceschini A., 1985, MNRAS 212, 857
%
\bibitem[1990]{cristiani:90}
Cristiani S., Vio R., 1990, A\&A 227, 385
%
\bibitem[1995]{cristiani:95}
Cristiani S., La Franca F., Andreani P., et al., 1995, A\&AS 112, 347
%
\bibitem[1976]{felten:76}
Felten J.E., 1976, ApJ 207, 700
%
\bibitem[1992]{goldschmidt:92}
Goldschmidt P., Miller L., La Franca F., Cristiani S.,
1992, MNRAS 256, 65
%
\bibitem[1978]{green:78}
Green R.F., Morrill M.E., 1978, PASP 90, 601
%
\bibitem[1993]{hewett:93}
Hewett P.C., Foltz C.B., Chaffee F.H., 1993, ApJ 406, L43
%
\bibitem[1995]{hewett:95}
Hewett P.C., Foltz C.B., Chaffee F.H., 1995, AJ 109, 1498
%
\bibitem[1992]{huchra:92}
Huchra J., Burg R., 1992, ApJ 393, 90
%
\bibitem[1996]{koehler:96}
K\"ohler T., 1996, Doctoral Thesis, Universit\"at Hamburg
%
\bibitem[1987]{marshall:87}
Marshall H., 1987, AJ 94, 628
%
\bibitem[1993]{miller:93}
Miller L., Goldschmidt P., La Franca F., Cristiani S., 
1993, Observational Cosmology, eds.\ G. Chincarini et al., 
ASP Conf.\ Proc.\ 51, 614
%
\bibitem[1996]{reimers:96}
Reimers D., K\"ohler T., Wisotzki L., 1996, A\&AS 115, 235 (Paper~II)
%
\bibitem[1973]{sandage:73}
Sandage A., 1973, ApJ 180, 687
%
\bibitem[1968]{schmidt:68}
Schmidt M., 1968, ApJ 151, 393
%
\bibitem[1983]{schmidt:83}
Schmidt M., Green R.F., 1983, ApJ 269, 352 (SG83)
%
\bibitem[1978]{spitzer:78}
Spitzer L., 1978, Physical Processes in the Interstellar Medium, Wiley\&Sons
%
\bibitem[1992]{stark:92}
Stark A.A., Gammie C.F., Wilson R.W., et al., 1992, ApJS 79, 77
%
\bibitem[1977]{terrell:77}
Terrell J., 1977, Am.\ J.\ Phys.\ 45, 869
%
\bibitem[1993]{veron:93}
V\'{e}ron-Cetty M.-P., V\'{e}ron P., 1993, A Catalogue of Quasars and
Active Galactic Nuclei, 6th ed., ESO Scientific Report 13
%
\bibitem[1985]{wampler:85}
Wampler E.J., Ponz D., 1985, ApJ 298, 448
%
\bibitem[1996]{wisotzki:96}
Wisotzki L., K\"ohler T., Groote D., Reimers D., 1996, A\&AS 115, 227 (Paper~I)
%
%
\end{thebibliography}
\end{document}